\documentclass[12pt, final, onecolumn ]{IEEEtran}

\usepackage{graphicx}
\usepackage{color}

 \usepackage{amsmath}
\usepackage{setspace}
\usepackage{enumerate}
\usepackage{verbatim}
\usepackage{amssymb}
\usepackage{amsbsy}
\usepackage{theorem}
\usepackage{url}

\usepackage[dvips]{epsfig}
\usepackage{float}

\def\urltilde{\kern -.15em\lower .7ex\hbox{\~{}}\kern .04em}
\def\urldot{\kern -.10em.\kern -.10em}

\def\urlhttp{http\kern -.10em\lower -.1ex\hbox{:}\kern -.12em\lower 0ex\hbox{/}\kern -.18em\lower 0ex\hbox{/}}

\theoremstyle{plain}
\newtheorem{Theorem}{Theorem}

\newtheorem{Proposition}{Proposition}
\newtheorem{Corollary}{Corollary}

{\theorembodyfont{\rmfamily} \newtheorem{Example}{Example}}
\newtheorem{Remark}{Remark}

\newcommand {\R}{\mathbb R}

\newcommand{\be}{\begin{equation}}
\newcommand{\ee}{\end{equation}}
\newcommand{\Int}{\operatorname{{\mathrm int}}}
\newcommand{\sgn}{\operatorname{{\mathrm sgn}}}

{\theorembodyfont{\rmfamily} }

\newcommand{\diag}{\operatorname{{\mathrm diag}}}

\newcommand{\myc}[1]{{\bf{#1}}}  

 \newcommand{\model}{MFALK}

\begin{document}
\title{A Deterministic
Mathematical Model for Bidirectional Excluded Flow with Langmuir~Kinetics\thanks{The research of MM and TT is partially supported by   research grants
from  the Israeli
Ministry of Science, Technology \& Space and the Binational Science Foundation. The research of MM is also
supported by a research grant form the Israeli Science Foundation.}
}
\author{Yoram Zarai, Michael Margaliot, and Tamir Tuller
\IEEEcompsocitemizethanks{
\IEEEcompsocthanksitem Y.
Zarai is with the School of Electrical Engineering, Tel-Aviv
University, Tel-Aviv 69978, Israel.
E-mail: yoramzar@mail.tau.ac.il
\IEEEcompsocthanksitem
M. Margaliot is with the School of Electrical Engineering and the Sagol School of Neuroscience, Tel-Aviv
University, Tel-Aviv 69978, Israel.
E-mail: michaelm@eng.tau.ac.il
\IEEEcompsocthanksitem
T. Tuller is with the department of Biomedical Engineering and the Sagol School of Neuroscience, Tel-Aviv
University, Tel-Aviv 69978, Israel.
E-mail: tamirtul@post.tau.ac.il

 }}

\maketitle
\doublespace


\begin{abstract}

 In many important cellular  processes, including
mRNA translation, gene transcription, phosphotransfer, and intracellular transport,
  biological ``particles''   move along some kind of ``tracks''.
 The    motion  of these particles  can be modeled  as a
 one-dimensional movement along an ordered   sequence  of sites. The biological particles (e.g.,
 ribosomes,  RNAPs,     phosphate groups, motor proteins)
 have volume and cannot surpass one another.
In some cases,  there is a preferred direction of
movement   along the track,  but in general the movement  may be two-directional,
 and furthermore the particles may attach or detach from various regions
along the tracks (e.g. ribosomes may drop off the mRNA molecule before reaching a stop codon).

We derive  a  new    deterministic mathematical model for such transport phenomena that may be interpreted as the dynamic
mean-field approximation of an important model from mechanical statistics called 
the asymmetric simple exclusion process~(ASEP) with Langmuir kinetics.
Using tools from the theory of monotone dynamical systems and
contraction theory we show that
the model admits a unique equilibrium, and that every solution
 converges to this  equilibrium.  This means that the occupancy
in all the sites along the lattice converges to a steady-state value
that depends on the parameters
but not on the initial conditions.
Furthermore, we show that the model  entrains (or phase locks)
to periodic excitations in any of its forward, backward, attachment, or detachment rates.

We demonstrate an application of this  phenomenological transport model
  for analyzing the effect of ribosome drop off in mRNA translation.
	One may perhaps expect that drop off from a jammed site may increase  the
	total   flow by reducing congestion.
Our results show that this is not true. Drop off has a substantial effect on the flow, yet always leads to a
reduction in the steady-state protein production rate.

 \end{abstract}


\section{Introduction}

Movement is essential for the functioning of cells. Cargoes like    organelles  and  vesicles
 must be carried  between different locations in the cells.
The information encoded in  DNA and mRNA molecules
must be decoded  by ``biological machines'' (RNA polymerases and ribosomes) that move along these molecules in a sequential order.

\begin{figure}[t]
 \begin{center}
\includegraphics[width= 12cm,height=16cm]{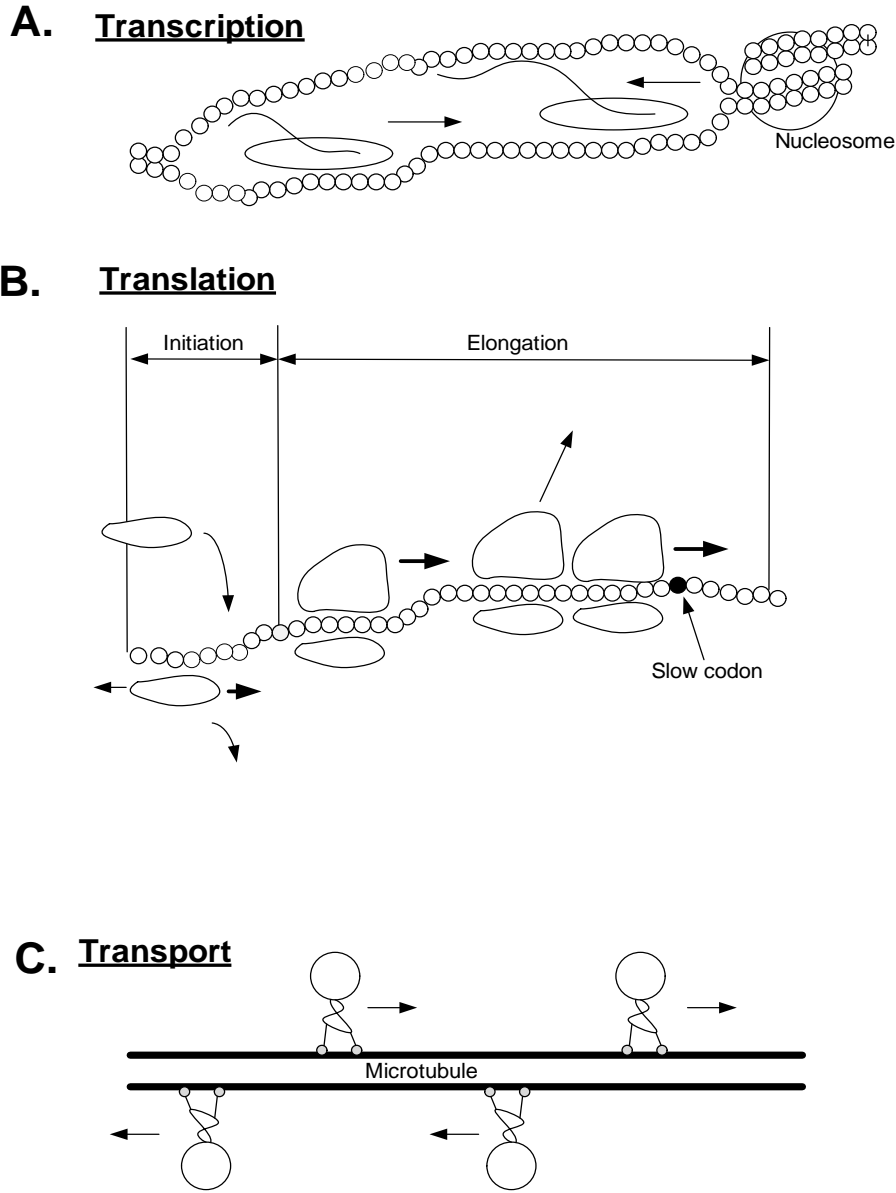}
\caption{Biological  processes  that can be studied using the model suggested in this paper. }\label{fig:biology}
 \end{center}
\end{figure}

Many of these important  biological transport processes  are    modeled as the movement of  particles along an ordered
 chain of sites.
In the context of intercellular transport, the particles are motor proteins and the chain models
actin   filaments or  microtubules. In transcription, the particles are RNAPs moving along the DNA molecule, and in translation
the particles are ribosomes moving along the mRNA molecule (see Figure~\ref{fig:biology}).

The movement in such processes may be  unidirectional, as in mRNA translation elongation, or bidirectional, as
in transcription or translation initiation. Indeed, the
  normal forward flow of the RNAP may be interrupted, due to transcription errors and various
  obstacles such as nucleosomes, in which case the RNAP tracks back a few nucleotides and then resumes its normal forward flow~\cite{backtracking_rnap, rnap_backtrack,structual_rnap,nascent_tran_vis}. Translation initiation in eukaryotes usually includes diffusion from the 5'end of the transcript towards the start codon~\cite{Alberts2002}. This diffusion process is believed to be bi-directional, but with a preference to the 5'$\to$3' direction.  The movement of
	motor proteins like  kinesin and dynein along  microtubules is  typically unidirectional, but can be two-directional as well \cite{Alberts2002}.

To increase efficiency,  many particles  may move simultaneously  along the same track thus pipelining the production process. For example,
to increase translation  efficiency, a  number of ribosomes may act simultaneously as polymerases on the same
 mRNA molecule \cite{Yonath2012,Arava2003}.

The moving  biological particles have volume and usually cannot overtake a particle in front of them. This means that
a slowly moving particle  may lead to the formation of a traffic jam behind it.
For example,
  Leduc et al.~\cite{motor_jams_2012} have studied
 Kip3, a yeast kinesin-8 family motor, and
 demonstrated that motor protein traffic  jams can exist, given the right conditions. Other studies have
 suggested that traffic jams of RNAP [ribosomes] may evolve
 during transcription [translation]~\cite{Arava2003,Klumpp2009,Chu2014}.

  In some of these biological transport
	processes, the biological machines may either attach or detach at various sites  along the tracks. For example,
    ribosomes may detach from the mRNA molecule before reaching the stop codon due to traffic ``jams'' and ribosome-ribosome interactions or due to depletion in the concentration of tRNAs~\cite{zhang2010global,Kurland1992,Sin2016}.
Also, it is known that
  kinesin-family motor proteins are more susceptible to dissociation when their pathway is blocked~\cite{dixit_2008,telley_2009}.

Defects in these transport processes
may  lead to severe diseases or may even be lethal. For example,~\cite{motor_rev_2003}
lists the implications  of malfunctions of
protein motors in disease and developmental defects.

Developing a better understanding of these dynamical
biological processes by combining
 mathematical modeling and biological experiments
will have far reaching implications
to basic science in  fields  such as  molecular evolution and functional genomics,
as well as applications in synthetic biology, biotechnology,  human health, and more.
Mathematical or computational modeling is especially important in developing approaches for
 manipulating and controlling these processes, e.g. in order to optimize various goals in  biotechnology.

A standard   model for such transport processes  is the \emph{asymmetric simple exclusion process}~(ASEP) \cite{Shaw2003,TASEP_tutorial_2011}. This is a stochastic model  describing   particles that hop
 along an ordered lattice of sites. Each site
 can be either empty or occupied by a single particle, and a
 particle can only hop to an empty site. This ``simple exclusion principle''
 represents the fact that the particles have volume and cannot overtake one another. Simple exclusion generates
     an  indirect coupling   between the particles.
	In particular,
 traffic jams
may develop behind  a slow-moving particle.

In ASEP, a particle may hop  to any of the two neighboring sites (but only if they are free).
Typically, a particle can attach the lattice in one of its ends and detach from the other end.
When particles can also attach or detach at internal sites  along the lattice, the
model is referred  to as ASEP with \emph{Langmuir kinetics}.
  In the special case where  the  hops are unidirectional, ASEP is sometimes referred to as the \emph{totally asymmetric simple exclusion process}~(TASEP). A
TASEP-like system
 with Langmuir kinetics has been used to  model  limit order markets in~\cite{willmann2002exact}, and  is often
used in modeling molecular motor traffic~\cite{PhysRevLett.90.086601,PhysRevE.70.046101,Lipowsky200553,lipowsky2001random,1751-8121-45-18-185002}.
More generally,
ASEP has become a fundamental model in non-equilibrium statistical mechanics, and has been applied to model numerous natural and artificial  processes including traffic and pedestrian flow, the movement of ants,  evacuation dynamics, and more~\cite{TASEP_book}.

In this paper, we introduce
a  deterministic  mathematical model that may be interpreted as the dynamic  \emph{mean-field approximation of
  ASEP with Langmuir kinetics}~(\model). We analyze the~{\model} using tools from systems and control theory.
  In particular, we apply some recent developments in contraction theory to prove
  that the model is globally asymptotically stable, and that it entrains
  to periodic excitations in the   transition/attachment/detachment rates.
	In other words, if these rates change periodically in time with some
	common period~$T$ then all the state-variables in the~{\model} converge to a periodic solution with period~$T$.
	This is important because many biological processes are excited by periodic signals (e.g. the 24h solar day
	or the periodic cell-division  process),
	and proper functioning  requires phase-locking or entrainment to these excitations.

Our work is motivated  by the analysis of a model for mRNA translation called the \emph{ribosome flow model~(RFM)}~\cite{reuveni}.
This is    the mean-field approximation of the \emph{unidirectional}
  TASEP    \emph{without} Langmuir kinetics (see, e.g.,~\cite[section 4.9.7]{TASEP_book} and \cite[p. R345]{solvers_guide}).
	Recently,
the RFM
 has been   studied extensively using tools from systems and control theory~\cite{HRFM_steady_state,zarai_infi,RFM_stability, RFM_feedback, RFM_entrain, RFM_concave,RFM_sense, RFMR,rfm_control}.
	The analysis is motivated by    implications to many important biological questions. For  example,
			the
	sensitivity of the protein production rate  to the initiation and elongation rates along the mRNA molecule~\cite{RFM_sense},
	maximization of protein production rate~\cite{RFM_concave},  the effect of ribosome recycling~\cite{RFM_feedback,RFMR},
	and the consequences  of competition for ribosomes
	on large-scale simultaneous mRNA translation in the cell~\cite{RFMNP}
	(see also~\cite{Heldt20150107,Algar:cdc2014} for some related models).

	The {\model} presented here
 is  much more general than the~RFM, and can  thus be used to model and analyze many transport phenomena, including all
 the biological processes mentioned  above, that cannot be  captured using the~RFM. We demonstrate this by using
 the {\model}
to model and analyze mRNA translation  with \emph{ribosome  drop off} - a feature     that cannot be modeled using the~RFM.
 
 Ribosome drop off is a fundamental phenomena that has received considerable attention
 (see, e.g., \cite{Sin2016,Keiler1996,Keiler2015,Zaher2011,Chadani2010,Subramaniam2014,Gilchrist2006,Kurland1992,Jorgensen1990,Hooper2000}). 
 In many cases, ribosome drop off  is deleterious to the cell since translation is the most energetically consuming process in the cell and,  furthermore,  drop off yields    truncated, non-functional proteins. Thus,  transcripts undergo selection to minimize  drop off
	or its energetic cost \cite{Zafrir2016,Tuller2015,Subramaniam2014,Gilchrist2006,Kurland1992,Hooper2000}. 
There are various hypotheses on the biological   advantages  of ribosome drop off. For example,
Zaher and Green~\cite{Zaher2009} have suggested that ribosome drop off   
  is related to proof reading. 
	One may perhaps expect that another advantage is that
	drop off from a jammed site may increase  the
	total   flow by reducing congestion.
Our results using analysis of the {\model} show that this is not true. Drop off has a substantial effect on the flow, yet it always leads to a
reduction in the steady-state protein production rate.


The remainder of this paper is organized as follows. The next section describes the new mathematical model.
 Section~\ref{sec:main_res} presents our main analysis
results. Section~\ref{sec:application}
describes the application of the {\model} to model mRNA translation with ribosome drop off.
The final section concludes and describes
 possible directions for further research. To streamline the presentation, all the proofs are placed in the Appendix.

\section{The  model}\label{sec:rfm+}
The {\model} is a set of~$n$ first-order nonlinear differential equations, where $n$ denotes the number of compartments or sites along the ``track". Each site is associated with  a state variable~$x_i(t)\in[0,1]$ describing the normalized ``level of occupancy'' at site~$i$ at time $t$, with~$x_i(t)=0$ [$x_i(t)=1$] representing that site~$i$ is completely free [full] at time~$t$.
Since~$x_i(t)\in[0,1]$ for all~$t$, it may also be interpreted as the probability that site~$i$ is occupied at time~$t$.

The {\model} contains four sets of non-negative parameters:
\begin{itemize}
\item $\lambda_i$, $i=0,\dots,n$, controls  the forward transition rate from site $i$ to site $i+1$,
\item $\gamma_i$, $i=0,\dots,n$, controls the backward transition rate from site $i+1$ to site $i$,
\item $\beta_i$, $i=1,\dots,n$, controls the attachment rate to site $i$,
\item $\alpha_i$, $i=1,\dots,n$, controls the detachment rate from site $i$,
\end{itemize}
where we  arbitrarily refer to left-to-right flow along the chain as forward flow, and to flow in the other direction as backward flow.


The dynamical equations describing the {\model}  are:
\begin{align}\label{eq:rfm+}
\dot{x}_1 &=\lambda_0(1-x_1)+\gamma_1 x_2 (1-x_1)+\beta_1(1-x_1) -\lambda_1 x_1(1-x_2)-\gamma_0 x_1-\alpha_1x_1, \nonumber \\
\dot{x}_2 &= \lambda_{1}x_{1}(1-x_2)+\gamma_2 x_{3}(1-x_2)+\beta_2(1-x_2)
  -\lambda_2 x_2(1-x_{3})-\gamma_{1} x_2 (1-x_{ 1})-\alpha_2 x_2,  \nonumber \\
&\vdots\nonumber\\
\dot{x}_{n-1} &=  \lambda_{n-2}x_{n-2}(1-x_{n-1})+\gamma_{n-1} x_{n }(1-x_{n-1})+\beta_{n-1}(1-x_{n-1})  -\lambda_{n-1} x_{n-1}(1-x_{n })
\nonumber \\ &-\gamma_{n-2} x_{n-1} (1-x_{n-2}) -\alpha_{n-1} x_{n-1},   \nonumber \\
\dot{x}_n &= \lambda_{n-1} x_{n-1}(1-x_n) + \gamma_n (1-x_n)+\beta_n(1-x_n) -\lambda_n x_n - \gamma_{n-1} x_n
(1-x_{n-1})-\alpha_n x_n.
\end{align}

\begin{figure*}[t]
\centering
\scalebox{1.0}{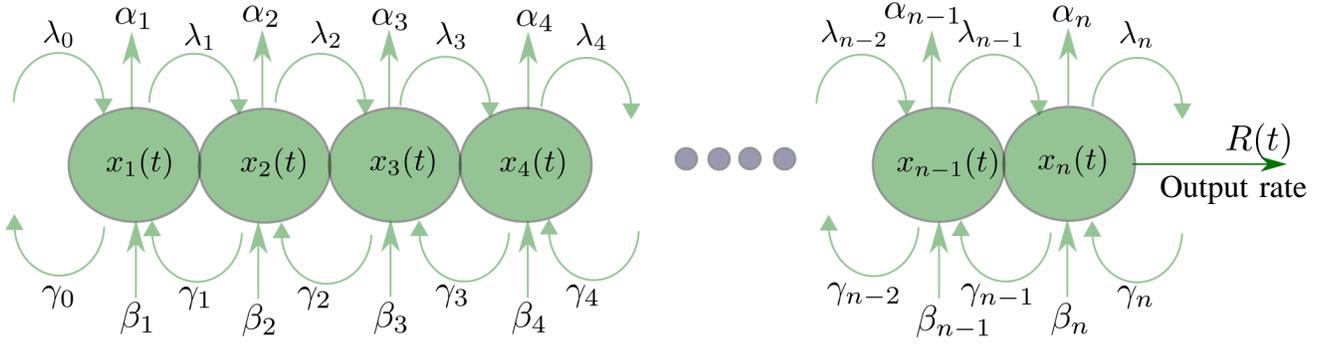}
\caption{Topology of the {\model}.}\label{fig:rfm+}
\end{figure*}

To explain these equations, consider for example 
the equation for the change in the occupancy in site~$2$, namely,
\[
           \dot{x}_2 =   \lambda_{ 1}x_{ 1}(1-x_2)+\gamma_2 x_{3}(1-x_2)+\beta_2(1-x_2) -
  \lambda_2 x_2(1-x_{3})-\gamma_{ 1} x_2 (1-x_{ 1})-\alpha_2 x_2.
\]
The term~$ \lambda_{ 1}x_{ 1}(1-x_2)$ represents the flow from site~$1$ to site~$2$.
This increases with the occupancy in site~$1$, and decreases with the occupancy in site~$2$.
In particular, this term becomes zero when~$x_2=1$, i.e. when site~$2$ is completely full.
This is a  ``soft''  version of the hard exclusion principle in ASEP:
the effective entry rate into a site decreases as it becomes fuller.
Note that the  constant~$\lambda_1\geq 0$ describes the maximal possible transition rate
 from site~$1$ to site~$2$.
Similarly, the term~$
  \lambda_2 x_2(1-x_{3})$ represents the flow from site~$2$ to site~$3$.
The term~$\gamma_2 x_{3}(1-x_2)$ [$\gamma_{ 1} x_2 (1-x_{ 1})$]
represents the backward flow from site~$3$ to site~$2$ [site~$2$ to site~$1$]. Note that these terms also
model    soft exclusion.
The term~$\beta_2(1-x_2)$   represents   attachment   of particles from the environment to
 site~$2$, whereas~$\alpha_2 x_2$ represents detachment of particles  from site~$2$ to the environment (see Fig.~\ref{fig:rfm+}).




The {\model} is  a \emph{compartmental model}~\cite{comp_surv_1993,sandberg78}, as
every  state-variable  describes
  the occupancy in a compartment   (e.g., a site along the the mRNA, gene, microtubule),
	and the dynamical
equations describe the flow between these compartments and the environment.
 Compartmental models play an important role
 in pharmacokinetics, enzyme kinetics, basic nutritional processes,
cellular growth, and  pathological processes, such as tumourigenesis
and atherosclerosis (see, e.g.,~\cite{comp_surv_1993,mrt_formulas} and the references therein).
More specifically,  the {\model}
is a nonlinear tridiagonal  compartmental model, as every~$\dot x_i$  directly
 depends    on~$x_{i-1},x_i$, and~$x_{i+1}$ only.

Note also that
\begin{align}\label{eq:sumdot}
			\sum_{i=1}^n\dot x_i&=\lambda_0 (1-x_1)-\gamma_0 x_1 +\beta_1(1-x_1) -\alpha_1 x_1\nonumber \\&
			                    +\gamma_n (1-x_n)-\lambda_n x_n +\beta_n(1-x_n) -\alpha_n x_n\nonumber  \\&+
													 \sum_{i=2}^{n-1} (\beta_i(1-x_i)-\alpha_i x_i).
\end{align}
The term on the right-hand side of the first [second] line here represents the change in~$x_0$ [$x_n$] due to the flow  between
the environment and site~$1$ [site~$n$], whereas the term on the third line represents the flow between internal sites and the environment.


The \emph{output rate} from site~$n$ at time~$t$
is the total flow from this site
to the   environment:
\begin{align}\label{eq:defrr}
R(t):&=(\lambda_n   +\alpha_n) x_n (t) - (\gamma_n +\beta_n) (1-x_n(t))  .
\end{align}
Note that~$R(t)$   may be positive, zero,  or negative.

In the particular case where~$\alpha_i =\beta_i=\gamma_i=0$ for all~$i$
the {\model} becomes the RFM, i.e. the dynamic
mean-field approximation of the unidirectional~TASEP with open boundary conditions
and without Langmuir kinetics.

Let~$x(t,a)$ denote the solution of~\eqref{eq:rfm+}
at time~$t \ge 0$ for the initial
condition~$x(0)=a$. Since the  state-variables correspond to normalized occupancy levels,
  we always assume that~$a$ belongs to the  closed $n$-dimensional
  unit cube:
\[
           C^n:=\{x \in \R^n: x_i \in [0,1] , i=1,\dots,n\}.
\]
Let~$\Int(C^n)$ denote the interior of~$C^n$, and let $\partial C^n$ denote the boundary of $C^n$.
The next section analyzes the {\model} defined in~\eqref{eq:rfm+}.

\section{Main Results}\label{sec:main_res}
\subsection{Invariance and persistence}
It is straightforward   to show that~$C^n$ is an invariant set for the dynamics of the~{\model}, that is,
if~$a\in C^n$ then~$x(t,a)\in C^n$ for  all~$t\geq0$.
The following result shows that a stronger property holds. Recall that all the proofs are placed in the Appendix.
For notational convenience, let~$\alpha_0:=0$,~$\gamma_0:=0$, $\alpha_{n+1}:=0$, and~$\beta_{n+1}:=0$.

\begin{Proposition}\label{prop:rfm+_inv1}
Suppose that at least one of the following two conditions holds:
\be
  \lambda_{i }+\beta_{i+1} >0,\quad \text{for all } i\in \{ 0,\dots, n\} ,\label{eq:cond1}
\ee
or
\be
\gamma_{i}+\alpha_{i+1} >0,\quad \text{for all } i \in \{0,\dots, n\}. \label{eq:cond2}
\ee
Then for any~$\tau>0$ there exists~$d=d(\tau)\in(0,1/2]$ such that
\be\label{eq:inter}
d   \leq x_i(t+\tau,a)  \leq 1-d,
\ee
for all $ a\in C^n$, all $i\in\{1,\dots, n\} $, and all $t\geq 0$.
\end{Proposition}

This means in particular  that trajectories that emanate
  from the boundary of~$C^n$ ``immediately'' enter~$C^n$.
	This result is useful   because as we will  see below
	on the boundary of~$C^n$ the {\model}
	looses some important  properties.
	For example, the Jacobian matrix of the dynamics~\eqref{eq:rfm+} is irreducible on~$\Int(C^n)$,
	but becomes reducible on some points on the boundary of~$C^n$.

\subsection{Contraction}
Differential analysis and in particular
contraction theory  proved to be a powerful tool for analyzing nonlinear dynamical systems. In a contractive system, trajectories that emanate from different initial conditions contract to each other at an exponential rate~\cite{LOHMILLER1998683,entrain2011,sontag_contraction_tutorial}.
Let~$|\cdot|_1:\R^n\to\R_+$ denote the~$L_1$ norm, i.e. for $z\in\R^n$,~$|z|_1=|z_1|+\dots+|z_n|$.

\begin{Proposition}\label{prop:rfm+_cont}
Let
\[
 \eta:=\max\{-\lambda_0-\gamma_0-\alpha_1-\beta_1, -\alpha_2-\beta_2,  \dots,  - \alpha_{n-1}-\beta_{n-1} ,- \lambda_n-\gamma_n-\alpha_n-\beta_n \}.
\]
Note that~$\eta \leq0$.
For any~$a,b\in C^n$ and any~$t\geq 0$,
\be\label{eq:conteta}
|x(t,a)-x(t,b)|_1\leq \exp(\eta t)|a-b|_1.
\ee
\end{Proposition}

This implies that the~$L_1$ distance between any two
trajectories contracts with the exponential rate~$\eta$.
Roughly speaking, this also means that increasing all the sums~$\alpha_i+\beta_i$, $i=1,\dots,n$,
 makes the system ``more contractive''.  Indeed,  these parameters have a direct stabilizing  effect on
the dynamics of site~$i$, whereas the   other parameters affect the site indirectly via the
coupling to the  two adjacent   sites.

When~$\eta=0$,~\eqref{eq:conteta} only  implies that the~$L_1$ distance between trajectories does not increase.
This property is not strong enough to prove the asymptotic properties described in the
subsections below.
Indeed, in this case it is possible that the {\model} will \emph{not} be contractive with respect to any fixed  norm.
Fortunately, a  certain generalization of contraction  turns out to hold in this case.

Consider  the time-varying dynamical system
\be\label{eq:tvsys}
\dot x(t)=f(t,x(t)),
\ee
 whose  trajectories   evolve
 on a compact and convex set~$\Omega\subset\R^n$. Let~$x(t,t_0,a)$ denote the
solution of~\eqref{eq:tvsys} at time~$t$ for the initial condition~$x(t_0)=a$.
System~\eqref{eq:tvsys}
is said to be   \emph{contractive after a small  overshoot}~(SO) \cite{cast15}
on~$\Omega$  w.r.t. a norm~$|\cdot|:\R^n\to\R_+$ if for any~$\varepsilon>0$
there exists~$\ell=\ell(\varepsilon)>0$
 such that
\[
|x(t ,t_0,a)-x(t,t_0 ,b)|\leq(1+\varepsilon)  \exp(-\ell t)|a-b|,
\]
for all $a,b\in\Omega$ and all~$t\geq t_0\geq 0$.
Intuitively speaking, this means contraction with an exponential rate,  but with  an arbitrarily
 small  overshoot of~$1+\varepsilon$.

 \begin{Proposition}\label{prop:weak_cont}
Suppose that
\be\label{eq:decup}
\lambda_i+ \gamma_i>0,\quad \text{for all } i \in \{ 1,\dots,n-1\},
\ee
and  that at least one of the  two conditions~\eqref{eq:cond1}, \eqref{eq:cond2} holds.
  Then the {\model} is SO on~$C^n$ w.r.t. the~$L_1$ norm, that is, for any~$\varepsilon>0$ there
	exists~$\ell=\ell(\varepsilon)>0$ such that
	\be\label{eq:sorfm}
|x(t ,a)-x(t ,b)|_1\leq(1+\varepsilon)  \exp(-\ell t)|a-b|_1,
\ee
for all $a,b\in C^n$ and all~$t\geq0$.
\end{Proposition}

Note that if~$\lambda_i+\gamma_i=0$ for some~$i\in\{1,\dots,n-1\}$, that is  $\lambda_i=\gamma_i=0$,
 then   the {\model}   decouples into two separate {\model}s: one containing  sites~$1,\dots,i$,
and the other containing sites~$i+1,\dots,n$.
Thus, assuming~\eqref{eq:decup} incurs no loss of generality.

There is an important difference between Propositions~\ref{prop:rfm+_cont} and~\ref{prop:weak_cont}.
If~$\eta<0$ then Proposition~\ref{prop:rfm+_cont} provides an explicit exponential contraction rate.
If~$\eta=0$ then Proposition~\ref{prop:weak_cont} can be used to deduce SO, but in
this result the contraction rate~$\ell$ depends on~$\varepsilon$ and is not given explicitly.

The contraction results above  imply that the {\model}
satisfies  several important
 asymptotic properties. These are described in the following subsections.

\subsection{Global asymptotic  stability}

Since the compact and convex set~$C^n$ is an invariant set of the dynamics, it contains an equilibrium point~$e$.
By Proposition~\ref{prop:rfm+_inv1}, $e\in\Int(C^n)$. Applying~\eqref{eq:sorfm} with~$b=e$
  yields the following result.

\begin{Corollary}\label{thm:rfm+_unique_e}
Suppose that the conditions in Proposition~\ref{prop:weak_cont}
 hold.
Then the {\model}
 admits a unique equilibrium point $e\in \Int(C^n)$ that is globally asymptotically stable, i.e. $\lim_{t\to\infty} x(t,a)=e$, for all~$a\in C^n$.
\end{Corollary}

This means that the rates determine a
unique distribution profile along the lattice, and that all trajectories
emanating from different initial conditions in~$C^n$ asymptotically converge to this  distribution.
 In addition, perturbations in the occupancy levels along the sites will not change this asymptotic behavior of the dynamics.
This also means   that various numerical solvers of ODEs will work well for the~{\model}
(see e.g.~\cite{Desoer_cont}).

\begin{figure}[H]
 \begin{center}
\includegraphics[width= 8cm,height=7cm]{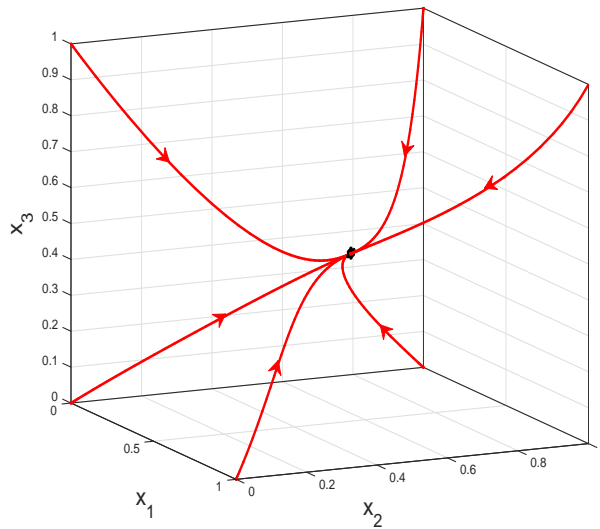}
\caption{Trajectories of the {\model} in Example~\ref{exa:tra} for six initial conditions in~$C^3$. }\label{fig:rfmp_ss}
 \end{center}
\end{figure}

\begin{Example}\label{exa:tra}
Fig.~\ref{fig:rfmp_ss} depicts the trajectories of a  {\model} with  $n=3$, $\lambda_0=1.0$, $\lambda_1=1.2$, $\lambda_2=0.8$, $\lambda_3=0.9$, $\gamma_i=\lambda_i-0.3$,  $i=0,\dots,3$, $\alpha_1=0$, $\alpha_2=0.1$, $\alpha_3=0$, $\beta_1=0$, $\beta_2=0.2$, $\beta_3=0$,
for  six
 initial conditions in~$C^n$. It may be seen
 that all trajectories converge to an equilibrium point~$e\in\Int(C^3)$.~\hfill{$\square$}
\end{Example}

The {\model}~\eqref{eq:rfm+} can be written as
\be\label{eq:rfm+_c}
\dot{x}_i = f_{i-1}(x)-f_i(x)+g_i(x_i),\quad  i=1,\dots,n,
\ee
where
\begin{align}\label{eq:defgg}
f_0(x)&:=\lambda_0(1-x_1)-\gamma_0 x_1, \nonumber  \\
f_i(x)&:=\lambda_i x_i (1-x_{i+1})-\gamma_i x_{i+1}(1-x_i), \quad  i=1,\dots,n-1, \nonumber\\
f_n(x)&:=\lambda_n x_n-\gamma_n(1-x_n), \nonumber \\
g_i(x_i)&:=\beta_i(1-x_i)-\alpha_i x_i,\quad i=1,\dots,n.
\end{align}

At steady-state, i.e. for $x=e$, the left-hand side of all  the equations in~\eqref{eq:rfm+_c} is zero, so
\be\label{eq:ffgg}
f_{i-1}(e)=f_i(e)-g_i(e_i), \quad  i=1,\dots,n.
\ee
Let~$v:=\begin{bmatrix}\alpha_1,\dots,\alpha_n,\beta_1,\dots,\beta_n,
\gamma_0,\dots,\gamma_n,\lambda_0,\dots,\lambda_n\end{bmatrix}'\in\R^{4n+2}_+$ denote   the
parameters of
the {\model}.
It follows from~\eqref{eq:ffgg} that if we multiply all these parameters
  by~$c>0$ then~$e$ will not change, that is,~$e(cv)=e(v)$.
Let
\be\label{eq:rfm+_R}
R:=(\lambda_n+\alpha_n) e_n - (\gamma_n+\beta_n)(1-e_n),
\ee
denote  the \emph{steady-state output rate}.
Then~$R(cv)=cR(v)$, for all~$c>0$,
that is, the steady-state production rate is homogeneous  of order one
w.r.t. the parameters.
By~\eqref{eq:ffgg},
\begin{align}\label{eq:rfm+_R_ss}
R&=f_n(e)-g_n(e_n) \nonumber \\
&=f_i(e)+\sum_{j=i+1}^{n-1} g_j(e_j), \quad  i=0,\dots,n-1.
\end{align}
This yields the following set of recursive   equations
relating the steady-state occupancy levels and the output rate in the~{\model}:
\begin{align}\label{eq:rfm+_e1}
e_n&=\frac{R+\gamma_n+\beta_n}{\lambda_n+\gamma_n+\beta_n+\alpha_n}, \nonumber \\
e_i&=\frac{R+\gamma_i e_{i+1}-\sum_{j=i+1}^{n-1} g_j(e_j)}{\lambda_i(1-e_{i+1})+\gamma_i e_{i+1}}, \quad  i=n-1,\dots,1,   \\
&\text{and also}\nonumber \\
e_1&=\frac{\lambda_0+\beta_1-R+\sum_{j=2}^{n-1} g_j(e_j)}{\lambda_0+\gamma_0+\beta_1+\alpha_1} \nonumber.
\end{align}
For a given~$v$,  this is a set of~$n+1$ equations in the~$n+1$ unknowns: $e_1,\dots,e_n,R$.

\begin{Example}
Consider the {\model} with dimension $n=2$. Then~\eqref{eq:rfm+_e1} becomes
\begin{align}\label{eq:14becom}
e_2 &= \: \frac{R+\gamma_2+\beta_2}{\lambda_2+\gamma_2+\alpha_2+\beta_2 }, \nonumber \\
e_1 &= \; \frac{R+\gamma_1 e_2 }{\lambda_1(1-e_2)+\gamma_1 e_2}, \\
&\text{and also}\nonumber \\
e_1 &= \; \frac{\lambda_0+\beta_1-R }{\lambda_0+\gamma_0+\beta_1+\alpha_1}\nonumber.
\end{align}
This   yields the polynomial equation~$a_2R^2+a_1 R+a_0=0$,
where
\begin{align*}
a_2&:=  \lambda_1-\gamma_1,  \nonumber\\
a_1&:=	 (\lambda_1-\gamma_1)(\gamma_2+\beta_2-\lambda_0-\beta_1)-\lambda_1 z_2-z_1 z_2 -z_1\gamma_1,  \nonumber \\
a_0&:=  (\lambda_0+\beta_1)\lambda_1(\lambda_2+\alpha_2) - (\gamma_0+\alpha_1)\gamma_1(\gamma_2+\beta_2) ,
\end{align*}
with $z_1:=\lambda_0+\gamma_0+\alpha_1+\beta_1 $ and $z_2:=\lambda_2+\gamma_2+\alpha_2+\beta_2 $.

Note that the polynomial equation admits several solutions~$R$, but only one solution corresponds to the
unique equilibrium point~$e\in C^2$. For example,
for~$\lambda_i=1$, $\gamma_i=2$, $\beta_i=3$, and~$\alpha_i=4$ for all~$i$
 the polynomial equation becomes
$-R^2-131 R-40=0$. This admits two solutions~$R_1=(-3 s-131)/2$ and~$R_2=(3 s-131)/2$, with~$s:=\sqrt{1889}$.
Substituting~$R_1$ in~\eqref{eq:14becom} yields~$e=[e_1 \; e_2 ]'$, with~$e_2<0$, so this is not a feasible solution.
Substituting~$R_2$ in~\eqref{eq:14becom} yields (all numbers are to four digit accuracy)~$e= \begin{bmatrix} 0.4305 & 0.4695 \end{bmatrix}'\in C^2$, which
 is the unique  feasible solution. Thus, the steady-state output rate is~$R_2=-0.3046$.~\hfill{$\square$}
\end{Example}

In general,~\eqref{eq:rfm+_e1}   can be transformed into a polynomial equation for~$R$. The  next result shows that
the degree
 of this polynomial equation grows  quickly with~$n$.
\begin{Proposition}\label{prop:brfm_degree}
Consider the {\model} with dimension~$n$ and with~$\lambda_i \ne \gamma_i$,  $\alpha_i \ne 0$, $\beta_i \ne 0$, for all~$i$. Then generically Eq.~\eqref{eq:rfm+_e1} may be written as $w(R)=0$, where $w(R)$ is a polynomial equation in~$R$ of degree~$1+\lfloor \frac{2^{n }}{3} \rfloor$,
 and with coefficients that are algebraic functions of the rates. 
\end{Proposition}

We note that this is exponential  increase in the degree of the polynomial equation is a feature of the {\model}
that does not take place in the RFM. Indeed, in the RFM the degree of the
polynomial equation for the steady-state production rate  grows linearly with~$n$.


Let~$\sgn(\cdot):\R\to\{-1,0,1\}$ denote the sign function, i.e.
\[
\sgn(y)=\begin{cases}1,& y>0,\\0,&y=0,\\-1,&y<0.\end{cases}
\]
An interesting question is what is~$\sgn(R)$. Indeed, if this is positive (negative)
then this means that there is a net  steady-state flow   from left to right (right to left).
The next subsection describes a special case where this question can be answered rigorously.
\subsubsection{Bidirectional flow with no  Langmuir kinetics}

In the case where $\beta_i=\alpha_i=0$, $i=1,\dots,n$, i.e. a system with no internal attachments and detachments,
Eq.~\eqref{eq:rfm+_R_ss} becomes
\be\label{eq:R_a=b=0}
R = f_i(e), \quad i=0,\dots,n.
\ee

\begin{Proposition}\label{prop:R_zero}
Consider the case where $\alpha_i=\beta_i=0$, $i=1,\dots,n$, and suppose that~\eqref{eq:decup} holds.
Then
\be\label{eq:sgneq}
\sgn(R)=\sgn\left(\prod_{i=0}^n \lambda_i - \prod_{i=0}^n \gamma_i\right).
\ee
In particular, if~$\prod_{i=0}^n \lambda_i = \prod_{i=0}^n \gamma_i$ then~$R=0$, and
\be\label{eq:ei_R0}
e_i=\frac{\prod_{j=0}^{i-1} \lambda_j}{\prod_{j=0}^{i-1} \lambda_j+\prod_{j=0}^{i-1} \gamma_j}=\frac{\prod_{j=i}^n \gamma_j}{\prod_{j=i}^n \gamma_j+\prod_{j=i}^n \lambda_j} ,  \quad i=1,\dots,n.
\ee
\end{Proposition}

Eq.~\eqref{eq:sgneq} means  that in the case of no  Langmuir kinetics
 the steady-state output from the right hand-side of the chain will be positive [negative] if the
   the product of the forward
rates is larger [smaller]
than  the product of the backward rates.
In      transcription and translation   the steady state flow from the right hand-side of the chain should always be positive,
but  in other cases, e.g.  transport along microtubules,  the steady state flow may  be either positive or negative.


\subsection{Entrainment}

Assume now that some  or all of the rates are time-varying periodic functions with the same period~$T$.
 This may be interpreted as a
periodic excitation of the system.
Many biological processes are affected by such excitations due for example to the periodic 24h solar day
 or the
periodic cell-cycle division process.
For example, translation elongation factors, tRNAs, translation and transcription initiation factors, ATP levels, and more may change in a periodic manner and affect   various rates that appear in the {\model}.

A natural question is will the state-variables of the {\model}
  converge to a periodic pattern with   period~$T$?
We will show that this is indeed so, i.e. the {\model} \emph{entrains} to a periodic excitation in the rates.
In order to understand what this means, consider
  a different setting, namely, using the {\model}
to model traffic flow. 
Then the  rates may correspond to 
traffic lights, changing in a periodic manner,  and the state-variables are the density of the moving particles (cars) along different sections of the road,
so entrainment corresponds to what is known as the ``green wave'' (see e.g.~\cite{0295-5075-102-2-28010} and the references therein).

We say that a function $f$ is $T$-periodic if $f(t+T)=f(t)$ for all $t$.
Assume that the $\lambda_i$s, $\gamma_i$s, $\alpha_i$s and~$\beta_i$s are uniformly bounded, non-negative,
time-varying functions satisfying:
\begin{itemize}
\item there exists a (minimal) $T>0$ such that all the $\lambda_i(t)$s, $\gamma_i(t)$s, $\alpha_i(t)$s, and $\beta_i(t)$s are $T$-periodic.
\item there exist~$c_1,c_2>0$ such that at least one of the following two conditions holds for all time~$t$
\begin{align}
 \lambda_{i }(t)+\beta_{i+1}(t)&>c_1 ,\quad i=0,\dots, n,\label{eq:cond1tv}\\
\gamma_{i}(t)+\alpha_{i+1}(t)&>c_2 ,\quad i=0,\dots, n \label{eq:cond2tv}.
\end{align}
\item there exists~$c_3>0$ such that
\begin{align}
 \lambda_{i }(t)+\gamma_{i+1}(t)&>c_3 ,\quad i=0,\dots, n. \label{eq:condd11}
\end{align}
\end{itemize}
We refer to this model as the \emph{Periodic {\model} (P{\model})}.

\begin{Theorem}\label{thm:rfm+_entr}
Consider the P{\model} with dimension $n$. There exists a unique function $\phi(\cdot):\R_+ \to \Int(C^n)$, that is $T$-periodic, and
for any~$a \in C^n$ the trajectory~$x(t,a)$ converges to~$\phi$ as~$t\to \infty$.
\end{Theorem}

Thus, the P{\model} \emph{entrains} (or phase-locks) to the periodic excitation in the parameters.
In particular, this means that the output rate $R(t)$
in~\eqref{eq:defrr} converges to the unique $T$-periodic function:
\[
(\lambda_n(t)+\gamma_n(t)+\beta_n(t)+\alpha_n(t))\phi_n(t)-\gamma_n(t)-\beta_n(t).
\]
Note that since a constant function is a periodic function for all $T\ge0$, Theorem~\ref{thm:rfm+_entr} implies that entrainment holds also in the particular case where a \emph{single} parameter is oscillating (with period $T>0$), while all other parameters are constant. Note also that Corollary~\ref{thm:rfm+_unique_e} follows from Theorem~\ref{thm:rfm+_entr}.

\begin{figure}[t]
 \begin{center}
\includegraphics[width= 9cm,height=8cm]{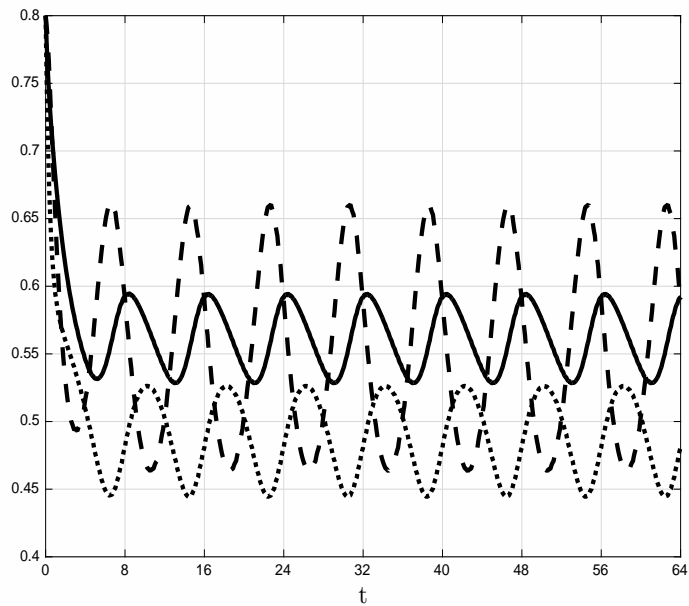}
\caption{State variables $x_1(t)$ [solid line]; $x_2(t)$ [dashed line]; and $x_3(t)$ [dotted line] as a function of $t$ in Example~\ref{exp:exp1}. Note that each state variable converges to a periodic function with a period $T=8$. }\label{fig:rfmp_x_per}
\end{center}
\end{figure}

\begin{Example}\label{exp:exp1}
Consider the {\model} with dimension $n=3$,  parameters:
$
\lambda_0(t) \equiv 1.0$,
$\lambda_1(t) \equiv 1.2$,
$\lambda_2(t) = 1+ 0.5\sin(\pi t/4 )$,
$\lambda_3(t) \equiv 0.9$,
$\gamma_0(t)  \equiv 0.4$,
$\gamma_1(t)  =0.4(1+\sin((\pi t/4) + 1/2))$,
$\gamma_2(t)  \equiv 0.25$,
$\gamma_3(t)   \equiv 0.45$,
$\alpha_1(t)\equiv 0$,
$\alpha_2(t)  \equiv 0.05$,
$\alpha_3(t)\equiv 0$,
$\beta_1(t)\equiv 0$,
$\beta_2(t) = 0.05(1+\sin((\pi t/2) + 1/4 ))$,
$\beta_3(t)\equiv 0$,
and   initial condition $x(0)=\begin{bmatrix} 0.8 & 0.8 & 0.8 \end{bmatrix}'$.
Note that all the rates here are periodic, with a minimal common period $T=8$.
 Fig.~\ref{fig:rfmp_x_per} depicts $x_i(t)$, $i=1,2,3$, as a function of $t$.
 It may be seen that each state variable converges to a periodic function with period~$T=8$.~\hfill{$\square$}
\end{Example}

\subsection{Strong Monotonicity}
Recall that a proper  cone~$K\subseteq \R^n$  defines a partial ordering in~$\R^n$ as follows.  For two vectors~$a,b \in \R^n$, we write~$a\leq b$ if~$(b-a) \in K$; $a<b$ if~$a\leq b$ and~$a \not =b$; and~$a \ll b$ if~$(b -a )\in \Int(K)$. The system~$\dot{y}=f(y)$ is called \emph{monotone} if~$a \leq b$ implies that~$y(t,a)\leq y(t,b)$ for all~$t \geq 0$. In other words, the flow preserves the partial ordering~\cite{hlsmith}. It is called \emph{strongly monotone} if~$a < b$ implies that~$y(t,a)\ll y(t,b)$ for all~$t > 0$.

From here on we consider  the  particular case where the cone is~$K:=\R^n_+$. Then~$a\leq b$ if~$a_i\leq b_i$ for all~$i$, and~$a \ll b$ if~$a_i <b_i$ for all~$i$. A system that is monotone with respect to this partial ordering  is called \emph{cooperative}.

\begin{Proposition}\label{prop:rfm+_mono}
						 For any~$a,b \in C^n$, with~$a \leq b$, the solutions of the {\model} satisfy
                \be\label{eq:abab}
                            x(t,a) \leq x(t,b), \quad \text{for all } t \geq 0.
                \ee
Furthermore, if~\eqref{eq:decup} holds
then
                \be\label{eq:strongabab}
                            x(t,a) \ll x(t,b), \quad \text{for all } t  > 0.
                \ee
\end{Proposition}

To explain this, consider two initial densities $a$ and~$b$ with~$a_i\leq b_i$ for all~$i$, that is, $b$
corresponds to a larger or equal density at each site. Then the trajectories~$x(t,a)$ and~$x(t,b)$ emanating from these initial conditions
continue to satisfy the same relationship between the densities, namely,
$x_i(t,a) \leq x_i(t,b)$,  for all~$i$ and  for all time~$t\geq0$.

The {\model} is thus
  a \emph{strongly cooperative tridiagonal system}~(SCTS) on $\Int(C^n)$. Some of the properties deduced above using    contraction theory
can also be deduced using this property~\cite{smillie}.
\begin{Remark}\label{rem:fint}
Suppose that we augment the {\model} into a model of~$n+1$ ODEs in~$n+1$ state-variables
by adding to it the equation
\begin{align*}
\dot x_{n+1}& =  - \lambda_0 (1-x_1)-\gamma_0 x_1 -\beta_1(1-x_1) +\alpha_1 x_1\nonumber \\&
			                    -\gamma_n (1-x_n) + \lambda_n x_n -\beta_n(1-x_n) +\alpha_n x_n\nonumber  \\&
													 -\sum_{i=2}^{n-1} (\beta_i(1-x_i)-\alpha_i x_i).
\end{align*}
that is,~$\dot x_{n+1}=-\sum_{i=1}^n \dot x_i$
(see~\eqref{eq:sumdot}). Let~$\tilde x$ denote the  vector of the~$n+1$ state-variables.
Clearly, this augmented model admits a first integral~$H(\tilde x(t)):=\sum_{i=1}^{n+1} \tilde x_i(t)$.
Also, for any initial condition in~$\tilde x(0) \in C^n \times \R_+$
 all the state-variables remain bounded, as the first~$n$ state-variables remain in~$C^n$
and~$\tilde x_{n+1}(t)=H(\tilde x(0))-\sum_{i=1}^{n } \tilde x_i(t)$ for all~$t\geq 0$.
It is straightforward to verify that the augmented system is a cooperative   system, and that
if~\eqref{eq:decup} holds then it is a~SCTS.  SCTS systems that admit a non-trivial
first integral have many desirable properties (see, e.g.~\cite{Mierc1991}).

\end{Remark}

\subsection{Effect of attachment and detachment}
One may perhaps expect that detachment from a jammed site may increase  the
	total   flow by reducing congestion.
The next result shows that this is not so.  Detachment always decreases the steady-state production rate~$R$.
Similarly, attachment always increases~$R$.

\begin{Proposition}\label{prop:brfm_sens}
Consider a  {\model} with dimension $n$. Suppose that the
conditions in Proposition~\ref{prop:weak_cont} hold. Then
$
\frac{\partial e_i}{\partial \alpha_j} < 0$,
and
$
\frac{\partial e_i}{\partial \beta_j} > 0$, for all~$i,j$.
Also,~$\frac{\partial R}{\partial \alpha_j} < 0$, and~$\frac{\partial R}{\partial \beta_j} > 0$
for all~$j=0,1,\dots,n-1$.
\end{Proposition}

\begin{figure}[t]
 \begin{center}
\includegraphics[width= 9cm,height=8cm]{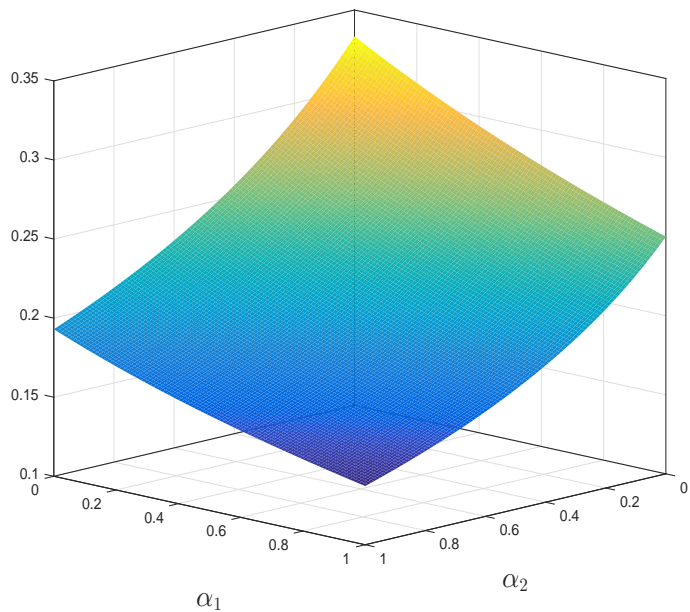}
\caption{$R$ as a function of $\alpha_1\in[0,1]$ and $\alpha_2\in[0,1]$ for  the
{\model} in Example~\ref{exa:detach3}.}\label{fig:brfm_n3_alpha}
\end{center}
\end{figure}

This means  that an increase in any of the detachment [attachment] rates   decreases [increases] the steady-state density in all the sites.  Also, an increase in any of the internal
detachment [attachment] rates   decreases [increases] the steady-state production rate.
The next example demonstrates  this. 
\begin{Example}\label{exa:detach3}
Consider the {\model} with $n=3$, $\lambda_i=1$, $\gamma_i=0$, $i=0,1,2,3$, $\beta_i=\alpha_3=0$, $i=1,2,3$. Fig.~\ref{fig:brfm_n3_alpha} depicts $R$ as a function of $\alpha_1\in[0,1]$ and $\alpha_2\in[0,1]$. It may be seen that $R$ decreases with both~$\alpha_1$ and~$\alpha_2$.~\hfill{$\square$}
\end{Example}

We note that the analytical results in Proposition~\ref{prop:brfm_sens}
agree well with the simulation results obtained using a TASEP model for translation that
included alternative initiation along the mRNA  and   ribosome drop-off~\cite{Zupniac14}.

The next section describes an application of the {\model} to a biological process.

\section{An application: modeling  mRNA translation  with ribosome drop off}\label{sec:application}


It is believed that during mRNA translation   ribosome movement is unidirectional
 from the 5' end to the 3' end, and that ribosomes do not
 enter in the middle of the coding regions. However, ribosomes
can detach  from various sites along
the mRNA  molecule
due for example to collisions between ribosomes.  This is known as ribosome drop off.

As mentioned in the introduction, ribosome drop off has been the topic of numerous  studies \cite{Sin2016,Keiler1996,Keiler2015,Zaher2011,Chadani2010,Subramaniam2014,Gilchrist2006,Kurland1992,Jorgensen1990,Hooper2000,Kurland1993}.
It was suggested that 
 in some cases ribosome drop off is important for proof reading \cite{Zaher2009}, and also
that  ribosome stalling/abortion plays a role in translational regulation (e.g. see~\cite{Shoemaker2010,Zupniac14}).

It is clear that ribosome  abortion has drawbacks. Indeed,
 translation is the most energetically consuming process in the cell,
 and abortion results in truncated, non-functional and possibly
deleterious proteins. It is believed that  transcripts undergo evolutionary selection to minimize abortion and/or its energetic cost \cite{Zafrir2016,Tuller2015,Subramaniam2014,Gilchrist2006,Kurland1992,Hooper2000}. Nevertheless,   there seems to be 
 a certain minimal abortion  rate     even in non-stressed
 conditions \cite{Sin2016,Kurland1993}. This  basal
 value was estimated (see more details below) to be  of the order or  $10^{- 4}-10^{- 3}$ abortion
events per codon in {\em E. coli}. 
In  other words, in every codon one out of $1,000-10,000$
decoding  ribosomes aborts.
  This value is   non-negligible. If we consider a drop-off rate of $ 4*10^{- 4}$ per codon   along
	a coding region of~$300$ codons (approximately the average coding region length for {\em E. coli})
	then on average, around $10$ out of every $100$ ribosomes will fail to complete the translation of the mRNA.

To model translation with ribosome drop off, we use the {\model}
with~$\gamma_i=0$ (i.e. no backwards motion) and~$\beta_i=0$ (i.e. no attachment to
 internal sites along the chain) for all~$i$.
Changing the values of the~$\alpha_i$s allows
to model and analyze the effect of ribosome drop off at different sites
along the mRNA molecule. We assume that
\be\label{eq:ahsa}
\lambda_i>0,\quad \text{for all }i,
\ee
 as otherwise the chain decouples into two smaller, disconnected chains.
Note that~\eqref{eq:ahsa}  implies that the conditions in Proposition~\ref{prop:weak_cont} hold,
so the model is  SO on~$C^n$ w.r.t. the~$L_1$ norm, and thus admits a unique globally asymptotically stable
equilibrium point~$e\in \Int(C^n)$.

\begin{figure}[t]
 \begin{center}
\includegraphics[width= 9cm,height=8cm]{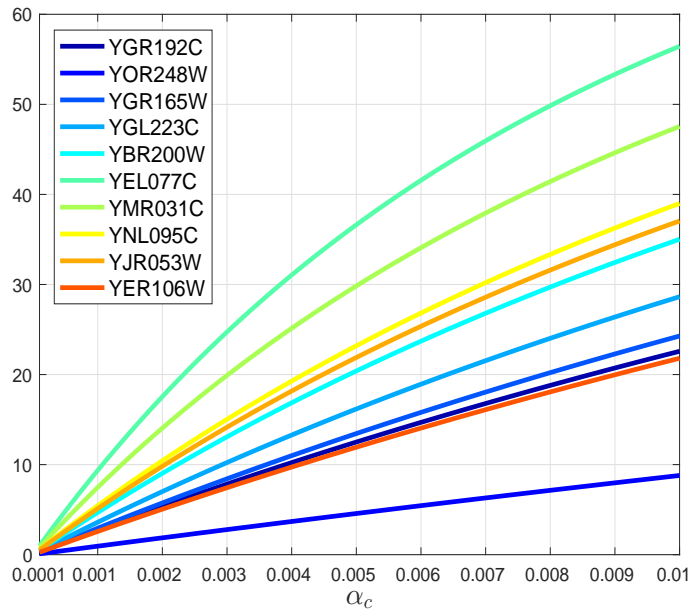}
\caption{ Reduction in steady-state mean density $\rho$ in percent
as a function of $\alpha_c\in[10^{-4}, 10^{-2}]$ for $10$ {\em S. cerevisiae} genes. }\label{fig:rfmp_e_drop}
\end{center}
\end{figure}

\begin{figure}[t]
 \begin{center}
\includegraphics[width= 9cm,height=8cm]{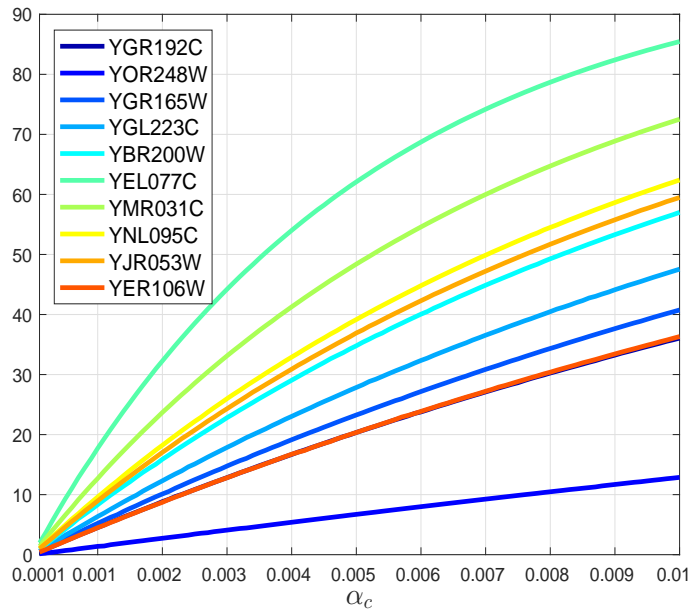}
\caption{ Reduction in steady-state output rate (production rate) $R$ in percent
as a function of $\alpha_c\in[10^{-4}, 10^{-2}]$ for $10$ {\em S. cerevisiae} genes. }\label{fig:rfmp_R_drop}
\end{center}
\end{figure}

We study the effect of ribosome drop off on the steady-state protein production rate and ribosome density
using real biological data. To this end, we
 considered $10$ {\em S. cerevisiae}
 genes (see Figures~\ref{fig:rfmp_e_drop} and~\ref{fig:rfmp_R_drop}) with various mRNA levels (all genes were sorted according to their mRNA levels and $10$ genes were uniformly sampled from the list).  Similarly to the approach used in \cite{reuveni}, we divided the mRNAs related to these genes to non-overlapping pieces. The first piece includes the first $9$ codons  that are   related to various 
 stages of initiation~\cite{Tuller2015}. The other pieces  include~$10$ non-overlapping codons each,
except for the  last one that includes between~$5$ and $15$ codons.

To model the translation dynamics in these mRNAs using {\model}, we model every piece of mRNA as a site.
We  estimated
the  elongation rates~$\lambda_i$ at  each site   using
   ribo-seq data for the
 codon decoding rates~\cite{Dana2014B},
 normalized so that the median elongation  rate of all {\em S. cerevisiae} mRNAs
becomes~$6.4$ codons per second \cite{Karpinets2006}. The site rate is~$(\text{site time})^{-1}$, where
 site time
 is the sum over the decoding times of all the codons in the piece of mRNA corresponding to this site.
These rates thus  depend on
various factors including  availability of tRNA molecules, amino acids, Aminoacyl tRNA synthetase activity and concentration, and local mRNA folding~\cite{Dana2014B,Alberts2002,Tuller2015}.

The initiation rate~$\lambda_0$  (that corresponds to the first piece)  was estimated based
 on the ribosome density per mRNA levels, as this value is expected to be approximately proportional to
the initiation rate when initiation is rate limiting \cite{reuveni,HRFM_steady_state}. Again we
applied a normalization that brings the median initiation rate of all {\em S. cerevisiae} mRNAs to be $0.8$ \cite{Chu2014}.

We analyzed the effect of uniform ribosome  drop off with a rate in the range of $10^{-5}$ to $10^{-3}$ per codon. This corresponds to $\alpha_1=\cdots=\alpha_n:=\alpha_c$, i.e., all the $\alpha_i$s are equal, and $\alpha_c$ denote their common value. Since we assumed $10$ codons per site, $\alpha_c$ values range from $10^{-4}$ to $10^{-2}$ (ten times the rate associated with  a  single codon). This   makes sense as in the {\model}
 the level of occupancy in a site is related to the probability to see a ribosome in this site.
  

Let
\[
\rho:=\frac{\sum_{i=1}^n e_i}{n},
\]
denote the steady-state mean ribosomal density. Figures~\ref{fig:rfmp_e_drop} and~\ref{fig:rfmp_R_drop}  depict
$\rho$ and~$R$  in our model as a function of $\alpha_c\in[10^{-4}, 10^{-2}]$. 
In these  figures   the genes in the legends are sorted according to their expression levels: the gene at the top (YGR192C) has the highest mRNA levels while the gene at the bottom (YER106W) has the lowest   levels.
  It may be seen that as the drop off (detachment) rate~$\alpha_c$
 increases from $10^{-4}$ to $10^{-2}$, $\rho$ decreases by about~$30\%$,
 and $R$ decreases by about~$50\%$. This demonstrate the significant ramifications
 that ribosomal drop off is expected to have on translation and the importance of  modeling drop off.

Note also that there is a strong variability in the effect of drop off on
 the different genes: for mRNAs with higher expression levels  (i.e. mRNAs with higher copy number in the cell) the drop off effect is weaker. 
It is possible that this  is related to stronger evolutionary selection for lower drop off rate in genes with higher mRNA levels. Indeed,  highly expressed genes  ``consume''
 more ribosomes 
(due to higher mRNA levels), so  a given
 (per-mRNA) drop off rate is expected to be more deleterious to the 
cell, and   a mutation which decreases the drop of rate in such genes has a  higher probability of fixation.

\section{Discussion}\label{sec:disc}

In many important
  processes  biological ``particles'' move along some kind of a  one-dimensional ``track''.
Examples include gene transcription and translation,   cellular transport, and more. The flow can be either bidirectional (as in the case of transcription) or unidirectional (as in the case of translation),
  with the possibility of both attachment and detachment of particles at different  sites
along the track.
For example, motor proteins like
kinesin and dynein that move along
a certain microtubule   may detach and attach to
   an  overlapping  microtubule.

To rigorously model and analyze such    processes, we introduced a new deterministic mathematical model that can be derived as the dynamic 
mean-field approximation of ASEP with Langmuir kinetics, called the {\model}. Our main results show that the {\model} is a monotone and contractive
dynamical system. This implies that it admits a globally asymptotically unique equilibrium point, and that it entrains to periodic excitations (with a common period $T>0$) in any of its rates, i.e. the densities along the chain, as well as the output rate, converge to unique period solutions with period $T$.

It is important to note that several
 known models   are special cases of the~{\model}. These include for example
the~RFM~\cite{reuveni}, the model used in~\cite{edri2014rna} for DNA transcription, and the model of phosphorelays
 in~\cite{phos_relays}.\footnote{Although in this
 model the occupancy levels are normalized differently.}


Topics for further research include the following. In the RFM, it has been
 shown  that the steady-state production rate is related to the maximal eigenvalue of a certain non-negative, symmetric tridiagonal matrix with elements that are functions of the RFM rates, i.e. the~$\lambda_i$s~\cite{RFM_concave}. This implies that the mapping $(\lambda_0,\dots,\lambda_n) \to R$ is strictly concave, and that sensitivity analysis of $R$ is an eigenvalue sensitivity problem~\cite{RFM_sense}. An interesting research topic is whether $R=R(\lambda_0,\dots,\lambda_n,\gamma_0,\dots,\gamma_n,\alpha_1,\dots,\alpha_n,\beta_1,\dots,\beta_n)$ in the {\model} can also be described  using such a linear-algebraic approach.

The application of  the  {\model}   to model  ribosome drop off suggests an interesting direction for further study, namely, 
 how to design genes that minimize  the drop off rate.


Another research direction
is motivated by the fact that many
  of the transport phenomena that can be modeled using the {\model} do not take place in isolation. For example, many
mRNA molecules are translated in parallel  in the cell.
Thus, a natural next step is to study networks of interconnected {\model}s.
Graph theory can be used to describe the interconnections between the various {\model}s   in the network.
In this context, ribosome drop off
may perhaps increase the total production rate in the entire system, as it allows ribosomes to detach
from slow sites, enter the pool of free ribosomes, and then attach to  the initiation
 sites of other, less crowded,   mRNA molecules. However, drop off still incurs the  biological ``cost''
associated to the synthesis  of
  a chain of amino-acids that is only a part of the desired protein.
The fact  that the {\model} is   contractive may prove useful in  analyzing
 networks of {\model}s, as
  there exist interesting
	results proving  the overall contractivity of a network
based on contractivity of
 the     subsystems and their couplings (see, e.g.~\cite{Arcak20111219,netwok_contractive}).

Another interesting topic for further research is studying the effect of
 controlled detachment rates on the formation of traffic jams.
Indeed, it is known that
  kinesin-family motor proteins are more susceptible to dissociation when their pathway is blocked~\cite{dixit_2008,telley_2009}.


\section*{Appendix: Proofs }
We begin by discussing
 some symmetry properties of the {\model}, as these will be useful in the proofs later on.
\subsection*{Symmetry}
The {\model} enjoys  two symmetries that will be useful later on.
First, let~$z_i(t):=1-x_{i}(t)$, $i=1,\dots,n$. In other words,~$z_i(t)$ is the amount of ``free space''
at site~$i$ at time~$t$.  Then using~\eqref{eq:rfm+}  yields
\begin{align}\label{eq:rfm_z_coor}
\dot{z}_1 &=    \gamma_0(1-z_1) +\lambda_1z_2(1-z_1) +\alpha_1(1-z_1) -\gamma_1z_1(1-z_2) -\lambda_0z_1-\beta_1 z_1 , \nonumber \\
\dot{z}_2 &=  \gamma_1 z_1(1-z_2) +\lambda_2 z_3(1-z_2) +\alpha_2(1-z_2) -\gamma_2 z_2(1-z_3) -\lambda_1 z_2(1-z_1)-\beta_2 z_2 , \nonumber  \\
&\vdots\nonumber\\
\dot{z}_n &= \gamma_{n-1} z_{n-1} (1-z_n) +\lambda_n  (1-z_n) +\alpha_n(1-z_n) -\gamma_n z_n
 -\lambda_{n-1} z_n (1-z_{n-1})-\beta_n z_n .
\end{align}
This is just the {\model}~\eqref{eq:rfm+}, but with the parameters permuted as follows:~$\lambda_k  \to \gamma_{k}$,
$\gamma_k \to \lambda_{k}$, $\beta_k \to \alpha_{k}$, and~$  \alpha_k \to \beta_{k}  $ for all~$k$.
The symmetry here follows from the fact that we can replace the roles of the
forward and backward flows in the {\model}.

Next, let~$y_i(t):=1-x_{n+1-i}(t)$, $i=1,\dots,n$. In other words,~$y_i(t)$ is the amount of ``free space''
at site~$n+1-i$ at time~$t$.  Then using~\eqref{eq:rfm+}  yields
\begin{align}\label{eq:rfm_ycoor}
\dot{y}_1 &=\lambda_n (1-y_1)+\gamma_{n-1} y_2 (1-y_1)+ \alpha_n
(1-y_1) -\lambda_{n-1} y_1(1-y_2)  -\gamma_n y_1 -\beta_n y_1, \nonumber \\
\dot{y}_2 &= \lambda_{n-1}y_{1}(1-y_2)+\gamma_{n-2} y_{3}(1-y_2)+\alpha_{n-1} (1-y_2)
  -\lambda_{n-2} y_2(1-y_{3})-\gamma_{n-1} y_2 (1-y_{ 1})-\beta_{n-1} y_2,  \nonumber \\
&\vdots\nonumber\\
\dot{y}_n &= \lambda_{1} y_{n-1}(1-y_n)  +\gamma_0(1-y_n) +\alpha_1 (1-y_n) -\lambda_0 y_n - \gamma_{1} y_n (1-y_{n-1})   -\beta_1 y_n.
\end{align}
This is just the {\model}~\eqref{eq:rfm+}, but with the parameters permuted as follows:~$\lambda_k  \to \lambda_{n-k}$,
$\gamma_k\to \gamma_{n-k}$, $\beta_k \to \alpha_{n+1-k}$, and~$  \alpha_k \to \beta_{n+1-k}  $ for all~$k$.
Note that~\eqref{eq:rfm_z_coor} is simply~\eqref{eq:rfm_ycoor} with the variable renaming $z_i \to y_{n+1-i}$, $i=1,\dots,n$.

Both symmetries are reminiscent of the \emph{particle-hole symmetry} in ASEP~\cite{solvers_guide,PhysRevE.71.011103}:
the basic idea is that the progression of a particle from left to right is also the progression of a hole  from right to left.

{\sl Proof of Proposition~\ref{prop:rfm+_inv1}.}
If~\eqref{eq:cond1} holds then the {\model}
satisfies property ({\bf BR}) in~\cite{RFM_entrain},
 and~\cite[Lemma~1]{RFM_entrain}  implies~\eqref{eq:inter}.
If~\eqref{eq:cond2} holds
then~\eqref{eq:rfm_z_coor} satisfies property ({\bf BR}) in~\cite{RFM_entrain},
and this implies~\eqref{eq:inter}.~\IEEEQED

{\sl Proof of Proposition~\ref{prop:rfm+_cont}}.
Write the {\model} as~$\dot x=f(x)$. A calculation shows  that the
  Jacobian matrix~$J(x):=\frac{\partial f}{\partial x}(x)$ satisfies~$J(x) = L(x)+ P $,
	where~$L(x)$ is   given in~\eqref{eq:rfm+_jac},
	and~$P$ is the diagonal matrix
	\be\label{eq:diagP}
	P=\diag( - \lambda_0-\gamma_0-\alpha_1-\beta_1, -\alpha_2-\beta_2,  \dots,   -\alpha_{n-1}-\beta_{n-1} , \lambda_n-\gamma_n-\alpha_n-\beta_n   ).
	\ee
	Note that~$L(x)$ is  tridiagonal  and Metzler (i.e, every off-diagonal entry is non-negative) for any~$x\in C^n$.

\begin{figure*}[!t]
\normalsize
\begingroup\makeatletter\def\f@size{11}\check@mathfonts
\be\label{eq:rfm+_jac}
L(x) =
\begin{bmatrix}
  -\lambda_1(1-x_2)-\gamma_1 x_2 & \lambda_ 1x_1 + \gamma_1(1-x_1) & \dots & 0 \\
\lambda_1(1-x_2)+\gamma_1 x_2 &  -\lambda_1 x_1 -\gamma_1(1-x_1)-\lambda_2(1-x_3)-\gamma_2x_3 & \dots & 0 \\
0 & \lambda_2(1-x_3)+\gamma_2 x_3 & \dots & 0 \\
& & \vdots & \\
0 & 0 & \dots & 0 \\
0 & 0 & \dots & \lambda_{n-1}x_{n-1}+\gamma_{n-1}(1-x_{n-1}) \\
0 & 0 & \dots &  -\lambda_{n-1}x_{n-1}-\gamma_{n-1}(1-x_{n-1}) \\
\end{bmatrix}
\ee
\endgroup
\hrulefill
\vspace*{4pt}
\end{figure*}

Recall that the matrix measure~$\mu_1:\R^{n\times n}\to\R$ induced by the~$L_1$ norm
is given by~$\mu_1(A)=\max\{c_1(A),\dots,c_n(A)\}$, where~$c_i(A)$ is the sum of the elements in column~$i$
of~$A$ with off-diagonal elements taken with absolute value~\cite{vid}. For the Jacobian~$J$ of the {\model},
$\mu_1(J(x))=\eta$ for all~$x\in C^n$.  It is well-known (see, e.g.,~\cite{sontag_contraction_tutorial}) that this
implies~\eqref{eq:conteta}.~\IEEEQED

 {\sl Proof of Proposition~\ref{prop:weak_cont}.}
For~$\zeta \in [0,1/2]$, let
\[
            C^n_\zeta:=\{x \in C^n:  \; \zeta  \leq x_i \leq  1-\zeta ,\; i=1,\dots,n\}.
\]
Note   that~$C^n_0=C^n$, and that~$ C^n_\zeta$
is a strict subcube of~$C^n$  for all~$\zeta \in (0,1/2] $.
By Proposition~\ref{prop:rfm+_inv1},
   for any~$\tau > 0$
 there exists~$\zeta=\zeta(\tau)\in (0,1/2)$,
with~$\zeta (\tau)\to 0$ as~$\tau \to 0 $,
such that
\be\label{eq:zinomeps}
            x(t +\tau,a) \in C^n_\zeta,\quad \text{for all } t\geq 0 \text{ and all }a\in C^n.
\ee
For any~$x\in C^n_\zeta$ every entry~$L_{ij}$ on the sub- and super-diagonal of~$L$ in~\eqref{eq:rfm+_jac}
satisfies~$   L_{ij}  \geq \zeta s $,
where~$s:=\min_{1\leq i \leq n-1}  \{ \lambda_i+\gamma_i \} > 0$.
Combining this with~\cite[Theorem~4]{RFM_entrain}, implies that for any~$\zeta\in(0,1/2]$ there exists
$\varepsilon=\varepsilon(\zeta)>0$, and a diagonal matrix~$D=\diag(1,q_1,q_1q_2,\dots,q_1q_2\dots q_{n-1})$,
with~$q_i=q_i(\varepsilon)>0$,
such that the {\model} is contractive on~$C^n_\zeta$
w.r.t. the    scaled~$L_1$ norm defined by~$|z|_{1,D}:=|Dz|_1$.
Furthermore, we can choose~$\varepsilon$ such that~$\varepsilon(\zeta) \to 0$ as~$\zeta\to 0$,
and~$D(\varepsilon )\to I$ as~$\varepsilon\to 0$.
Now Thm.~1 in~\cite{cast15} implies that the {\model}
is contractive  after a small overshoot and short transient~(SOST).
 Prop.~4 in~\cite{cast15} implies that for the {\model} SOST  is equivalent to~SO,
and this completes the proof.~\IEEEQED


{\sl Proof of Proposition~\ref{prop:brfm_degree}.}
We begin by recursively defining two sequences.
For all integers~$i\geq 1$, let
\begin{align}\label{eq:2seq}
					u_{i+1}&=1+\ell_1+\ell_2+\dots +\ell_i,  \nonumber \\
					\ell_{i+1}&=u_i+ \ell_1+\ell_2+\dots+\ell_{i-1}  .
\end{align}
with initial conditions~$u_0=u_1=1$, and~$\ell_0=0$, $\ell_1=1$.
 We claim that for~$k=0,1,\dots,n-1$ the steady-state density in site~$n-k$
  is generically the ratio of two polynomials in~$R$:
\be\label{eq:indu}
e_{n-k}=\frac{p_k(R)}{q_k(R)},\quad  \text{with }\deg(p_k(R))=u_k,\; \deg(q_k(R))=\ell_k.
\ee
 We  prove  this by induction on~$k$.
By~\eqref{eq:rfm+_e1}, $e_n=a R+b$, with~$a:=(\lambda_n+\gamma_n+\beta_n+\alpha_n )^{-1}$ and~$b:=(\gamma_n+\beta_n)a$,
and this proves~\eqref{eq:indu} for~$k=0$. Using~\eqref{eq:rfm+_e1} again  yields
\begin{align*}
e_{n-1}&=\frac{R+\gamma_{n-1} e_{n}}{\lambda_{n-1}(1-e_{n})+\gamma_{n-1} e_{n}}\\
&=\frac{R+\gamma_{n-1} (aR+b)}{\lambda_{n-1}+(\gamma_{n-1}-\lambda_{n-1}) (aR+b)},
\end{align*}
and this proves~\eqref{eq:indu} for~$k=1$. Now assume that there exists~$s\geq 2$
such that~\eqref{eq:indu} holds for~$k=0,1,\dots,s-1$.
By~\eqref{eq:rfm+_e1},
 \begin{align*}
e_{n-s}&=\frac{R+\gamma_{n-s} e_{n-s+1}-  g_{n-s+1}(e_{n-s+1})  - g_{n-s+2}(e_{n-s+2})  -\dots -g_{n-1}(e_{n-1})   }{\lambda_{n-s}(1-e_{n-s+1})+\gamma_{n-s}
 e_{n-s+1}},
\end{align*}
and applying~\eqref{eq:defgg} and  the induction hypothesis yields
 \begin{align*}
e_{n-s}&=\frac{R+\gamma_{n-s} \frac{p_{s-1}}{q_{s-1}} +(\beta_{n-s+1}+\alpha_{n-s+1}) \frac{p_{s-1}}{q_{s-1}}
+  (\beta_{n-s+2}+\alpha_{n-s+2}) \frac{p_{s-2}}{q_{s-2}} +\dots + (\beta_{n-1}+\alpha_{n-1})\frac{p_{1}}{q_{1}} +c }
{\lambda_{n-s}+ (\gamma_{n-s}-\lambda_{n-s}) \frac{p_{s-1}}{q_{s-1}}},
\end{align*}
where~$c:= -\beta_{n-s+1}-\dots-\beta_{n-1}$.
Multiplying the numerator and the denominator by~$q_1\dots q_{s-1}$  yields~$e_{n-s}= {p_s}/{q_s}$,
where\begin{align*}
\deg(p_s)&=\max\{1+\deg(q_1\dots q_{s-1}) , \deg(p_{s-1} q_1\dots q_{s-2}),\dots,\deg( p_1 q_2\dots q_{s-1}) \},\\
\deg(q_s)&=\max\{\deg(q_1\dots q_{s-1}) , \deg(p_{s-1} q_1\dots q_{s-2})\}.
\end{align*}
By the induction hypothesis,
\begin{align}\label{eq:pos}
\deg(p_s)&=\max\{1+\ell_1+\dots +\ell_{s-1}  ,  u_{s-1}+ \ell_1+\dots+ \ell_{s-2} ,\dots,u_1 +\ell_2+\dots+ \ell_{s-1} \},\nonumber \\
\deg(q_s)&=\max\{\ell_1+\dots +\ell_{s-1} ,u_{s-1}+ \ell_1+\dots+ \ell_{s-2}\}.
\end{align}

It is straightforward to prove that~\eqref{eq:2seq} implies that
\be\label{eq:ra}
\ell_i\leq u_i\leq \ell_i+1,\quad i=0,1,2,\dots.
\ee
Combining this with~\eqref{eq:pos} yields~$\deg(p_s) = 1+\ell_1+\dots +\ell_{s-1}$, and~$\deg(q_s) = u_{s-1}+ \ell_1+\dots+ \ell_{s-2}
$. Thus,~$\deg(p_s) = u_s$ and~$\deg(q_s)=\ell_s$, and this completes the inductive proof of~\eqref{eq:indu}.
In particular,~\eqref{eq:indu} yields
\be
e_{1}=\frac{p_{n-1}(R)}{q_{n-1}(R)},
\ee
with~$\deg(p_{n-1}(R))=u_{n-1},\; \deg(q_{n-1}(R))=\ell_{n-1}$.
Substituting this in the last equation of~\eqref{eq:rfm+_e1} yields
\[
v \frac{p_{n-1} }{q_{n-1} }= z-R+\sum_{j=2}^{n-1} g_j(e_j)  ,
\]
where~$v:=\lambda_0+\gamma_0+\beta_1+\alpha_1$, and~$z:=\lambda_0+\beta_1$.
Arguing as above shows that this is
a polynomial equation of the form~$w(R)=0$, with~$\deg(w)=1+\ell_1+\dots+\ell_{n-1}=u_n$.
It is straightforward to prove by induction that~\eqref{eq:2seq} implies that
\begin{align*}
u_k =1+\left \lfloor\frac{2^{k }}{3} \right \rfloor,\quad
\ell_k =\frac{ 2^k-(-1)^k }{3},
\end{align*}
(we note in passing that the latter sequence is  known as the  		Jacobsthal sequence~\cite{Sloane_seq}),
and this completes the proof of Proposition~\ref{prop:brfm_degree}.~\IEEEQED

{\sl Proof of Proposition~\ref{prop:R_zero}.}
We begin by proving that~$R>0$ implies that~$\prod_{i=0}^n \lambda_i>\prod_{i=0}^n \gamma_i $.
If~$R>0$ then~\eqref{eq:R_a=b=0} yields
\begin{align}\label{eq:R0_equ}
\lambda_0(1-e_1) &> \gamma_0 e_1, \nonumber \\
\lambda_i e_i (1-e_{i+1})&>\gamma_i e_{i+1}(1-e_i), \quad  i=1,\dots,n-1, \nonumber \\
\lambda_n e_n &> \gamma_n (1-e_n ).
\end{align}
Multiplying all these inequalities, and using the fact  that~$e\in\Int(C^n)$ yields
\be\label{eq:afii}
\prod_{i=0}^n \lambda_i>\prod_{i=0}^n \gamma_i.
\ee

To prove the converse implication, assume that~\eqref{eq:afii} holds.
 Multiplying both sides of this inequality by the strictly positive term~$\prod_{j=1}^n e_j (1-e_j)$  yields
 \[
 \prod_{i=0}^n a_i > \prod_{i=0}^n b_i,
 \]
where
  $a_0:=\lambda_0 (1-e_{1})$, $a_i:=\lambda_i e_i(1-e_{i+1})$, $i=1,\dots,n-1$, $a_n=\lambda_n e_n$, $b_0:=\gamma_0 e_{1}$, $b_i:=\gamma_i e_{i+1}(1-e_{i})$, $i=1,\dots,n-1$,  and~$b_n=\gamma_n (1-e_n)$.
 This means that~$a_\ell>b_\ell$ for some index~$\ell \in\{0,\dots,n\}$.
Since $R=a_\ell-b_\ell$ (see~\eqref{eq:R_a=b=0}), it follows that $R>0$. Summarizing, we showed that~$R>0$ if and only if~$\prod_{i=0}^n \lambda_i>\prod_{i=0}^n \gamma_i$. The proof that~$R<0$ if and only if~$\prod_{i=0}^n \lambda_i < \prod_{i=0}^n \gamma_i$ is similar.
This implies that~$R=0$ if and only if~$\prod_{i=0}^n \lambda_i =  \prod_{i=0}^n \gamma_i$.
This completes the proof of~\eqref{eq:sgneq}.

To prove~\eqref{eq:ei_R0}, note that~\eqref{eq:R_a=b=0} yields
\begin{align}\label{eq:ei_b=a=0}
e_n &=\frac{R+\gamma_n}{\lambda_n+\gamma_n}, \nonumber \\
e_i &= \frac{R+\gamma_i e_{i+1}}{\lambda_i(1-e_{i+1})+\gamma_i e_{i+1}}, \quad i=n-1,\dots,1, \nonumber  \\
 e_1 &= \frac{\lambda_0-R}{\lambda_0+\gamma_0}.
\end{align}
Substituting~$R=0$ completes the proof of  Prop.~\ref{prop:R_zero}.~\IEEEQED

{\sl Proof of Proposition~\ref{prop:rfm+_mono}.}
Since the  Jacobian~$J(x)$ of the {\model} 	 is Metzler (i.e, every off-diagonal entry is non-negative) for any~$x\in C^n$,
	   the {\model}
	is a  cooperative system~\cite{hlsmith}, and this yields~\eqref{eq:abab}.
	
	When~$\lambda_i+ \gamma_i>0$,~$i=1,\dots,n-1$,  the matrix~$L(x)$  and, therefore,~$J(x)$, is irreducible
	  for every~$x \in \Int(C^n)$, and combining this with Proposition~\ref{prop:rfm+_inv1}
		  implies~\eqref{eq:strongabab} (see, e.g.,~\cite[Ch.~4]{hlsmith}).~\IEEEQED

{\sl Proof of Theorem~\ref{thm:rfm+_entr}.}
The Jacobian of the~P{\model}
is~$J(t,x(t))=L(t,x(t))+P(t)$, with~$L$ given in~\eqref{eq:rfm+_jac},
and~$P$ is given in~\eqref{eq:diagP} (but now with time-varying rates).
Pick an initial time~$t_0\geq0$, and~$\tau_0>0$.
The stated conditions guarantee the existence of~$\zeta\in(0,1/2)$ such that~$x(t,t_0,a)\in C^n_\zeta$
for all~$t\geq t_0+\tau$ and all~$a\in C^n$. Also,~\cite[Thm.~4]{RFM_entrain}  implies that there exists a diagonally-scaled $L_1$
norm such that
the~P{\model} is contractive on~$ C^n_\zeta$ w.r.t. this norm. Now entrainment follows from known results on
contractive systems with a  periodic excitation (see, e.g.~\cite{entrain2011}).~\IEEEQED

 {\sl Proof of Proposition~\ref{prop:brfm_sens}.}
First, using Remark~\ref{rem:fint} and the argument used in the proof of~\cite[Prop.~4]{RFMNP}
shows  that all the derivatives in the statement of of Proposition~\ref{prop:brfm_sens} exist.

Given  a~{\model}, pick~$j \in \{1,\dots, n\}$ and consider
 the new {\model} obtained by changing~$\alpha_j$
to~$\tilde \alpha_j$, with~$\tilde \alpha_j>\alpha_j$, and all other rates unchanged.
Let~$\tilde e$, $\tilde R$ denote the steady-state density and production rate in the modified {\model}.
Seeking a contradiction, assume that
\be\label{eq:asmp1}
 \tilde e_n \geq e_n.
\ee
Then~\eqref{eq:rfm+_R} implies that
\be\label{eq:rrtep}
\tilde R \geq R,
\ee
and if~$j=n$ then~$\tilde R>R$.
By~\eqref{eq:rfm+_R_ss} with~$i=n-1$,  $R = \lambda_{n-1} e_{n-1}(1-e_n)-\gamma_{n-1} e_{n}(1- e_{n-1})$ and~$\tilde
R = \lambda_{n-1} \tilde e_{n-1}(1-\tilde e_n)-\gamma_{n-1} \tilde e_{n}(1- \tilde e_{n-1})$, and combining this with~\eqref{eq:asmp1} and~\eqref{eq:rrtep} yields
\be\label{eq:asmp2}
  \tilde e_{n-1} \geq e_{n-1}.
\ee
Now using~\eqref{eq:rfm+_R_ss} with $i=n-2$ yields~$\tilde e_{n-2} \geq e_{n-2}$,
and~$\tilde e_{n-2} > e_{n-2}$ if~$j=n-1$.
Proceeding in this way shows that
\begin{align}\label{eq:trew}
								\tilde e_{k} &\geq e_{k},\quad k=n,n-1,\dots,j,\\
									\tilde e_{k} &> e_{k},\quad k=j-1,j-2,\dots,1.
\end{align}
Combining  
 this with~\eqref{eq:rfm+_R_ss} with $i=0$ yields~$\tilde R<R$.
This contradicts~\eqref{eq:rrtep}, so
\be\label{eq:en_ineq}
 \tilde e_n > e_n.
\ee
Proceeding as above yields~$\tilde e_i>e_i$ for all~$i$, so~$\frac{\partial e_i}{\partial \alpha_j}< 0$ for all~$i,j$.
The proofs of all the other equations  in
 Prop.~\ref{prop:brfm_sens}  are very similar and therefore omitted.~\IEEEQED


\end{document}